\documentclass[pra,twocolumn,showpacs]{revtex4}
\usepackage{graphicx}
\usepackage{dcolumn}
\usepackage{amsmath}
\usepackage{bm}

\begin{document}

\preprint{cond-mat/010000}

\title{Collective excitations of trapped Bose condensates in the
energy and time domains}

\author{Dermot McPeake}
\email{d.mcpeake@am.qub.ac.uk}
\author{Halvor M\o ll Nilsen}
\email{halvor.nilsen@fi.uib.no}
\author{J F McCann}
\affiliation{Department of Applied Mathematics and Theoretical
Physics, Queen's University Belfast, Belfast BT7 1NN, Northern
Ireland }
 \affiliation{ Institute of Physics, University of
Bergen, All\'egaten 55, N-5007 Bergen, Norway}

\date{November 2001}.

\begin{abstract}
A time-dependent method for calculating the 
collective excitation
frequencies and densities of a trapped, inhomogeneous Bose-Einstein 
condensate with  circulation is presented. The results are compared with 
time-independent solutions of the
Bogoliubov-deGennes equations. The method is based on 
time-dependent linear-response theory combined 
with spectral analysis of
moments of the excitation modes of interest. The 
technique is
straightforward to apply, is extremely efficient in our
implementation with parallel FFT  methods, and produces highly
accurate results.  The method is suitable for general trap
geometries, condensate flows and condensates 
permeated with vortex structures. 
\end{abstract}


\maketitle

\section{Introduction}

The equations defining the  frequencies of single particle
excitations from a weakly-interacting dilute Bose gas at zero
temperature were introduced by Bogoliubov \cite{bog47} and
applied  to low-temperature Fermi
systems by de Gennes \cite{deg66}. The underlying assumption
is that temperatures are sufficiently low such that the condensate is not
significantly depleted and that the interparticle interactions 
are weak and the gas dilute. Under such constraints
the dynamics are dominated by single particle
excitations meaning that semiclassical approximations can be used
for the field excitations and the equations are  equivalent to
linear response theory \cite{pin60,edw96,dal99}. Trapped  
hydrostatic ultracold
gas clouds can be modeled to a high degree of precision under
these conditions. 
Even within this framework the equations must be solved numerically because
of the inhomogeneity and nonlinearity  of the condensate.
Determining the elementary excitations
is much more complex in
situations in which the  condensate is flowing or  
permeated with defects such as vortex arrays. 
The extraordinary coldness of the atom cloud means the
condensate dynamics occur on millisecond time scales and 
thus are accessible to measurement as they evolve.
Shape distortions,  created  by  manipulating the condensate
with external electromagnetic fields, 
 can be used to measure  mode 
frequencies \cite{jin96,and97,sta98}, wave-mixing \cite{den99}   and 
damping rates \cite{mar01}.
In this way elementary excitation  
frequencies at low temperatures 
can be measured to an accuracy within a few percent, and 
the agreement with theory for the low-lying modes 
 is astonishingly good \cite{dal99}. Precise measurements closer to the
critical temperature are much more uncertain, and indeed theoretical 
modeling is also  challenged by the influence of  complex pair
excitations  \cite{hut98}. The agreement between theory and 
experiment is much less 
satisfactory in this region.

The experimental and
theoretical study of condensates in the time domain 
mirrors experimental breakthroughs in ultrashort laser pulses
used to probe  ultrafast processes in chemistry and biology.
Dynamic simulations and experiments applied 
to Bose condensates have provided insight into  the
transition from superfluid flow to 
dissipation and drag \cite{ram99,win00} and in condensate 
formation and destruction. Trapped condensates are particularly
suited to spectral methods, specifically the split-operator
technique \cite{tah84} combined with the FFT method 
\cite{fei82}. We report results in which efficient parallel
computing techniques have been implemented within the spectral 
method to determine the elementary excitation frequencies 
and densities. The method is ideally suited to the
analysis of elementary excitations 
in complex flows and in the presence of defects. The method
also allows the excitation mechanism of experiments 
to be modeled realistically  and thus help
devise schemes which produce optimal excitation of
certain modes or superposition of modes leading to squeezed and
entangled states \cite{orz01}, and observe collective excitations
of quadrupoles giving evidence of superfluid motion \cite{mar00,gue99}.

The paper is structured as follows: In section~\ref{sec:QFT} the
units and notation are described,  and the 
field equations are introduced.  
The case of excitation in the
presence of condensate circulation is discussed. 
The equations describing excitations in a cylindrically symmetric trap and
quadrupole excitations in an asymmetric trap are given. In section~\ref{sec:tdlr} the time-dependent method is formulated. Finally, in sections~\ref{sec:res} and~\ref{sec:conc}, results are compared betwen the two methods and some general conclusions are given on the relative merits of each method.
\section{Time-independent quantized field equations}\label{sec:QFT}
\label{sec:level1}

\begin{table}
\begin{tabular}{|r r |r|r|}
\hline
Interaction & Strength & \multicolumn{2}{c|}{Chemical potential ($\mu$)} \\
\multicolumn{2}{|c|}{$C$} & \multicolumn{2}{c|}{$\kappa=0 \ \ \ \ \ \ \kappa=1$} \\
\hline
0    & &       1.000000  &   2.000000      \\
10   & &         1.620434  &   2.368791      \\
50   & &         3.020229  &   3.489876      \\
100  & &         4.143006  &   4.513850      \\
250  & &         6.415956  &   6.685521      \\
500  & &         9.003106  &   9.213175      \\
1000 & &        12.678320  &  12.840724      \\
\hline
\end{tabular}
\caption{Values of the chemical potential ($\mu$) of the
$\kappa=0$ and $\kappa=1$ states for the isotropic 2D trap as a
function of interaction strength, $C$, in  units of $\hbar
\omega_0$}
\label{chempot}
\end{table}

\begin{table}
\begin{tabular}{|c|c|c|c|cc|}
\hline
 & \multicolumn{2}{c}{$\kappa=0$}&\multicolumn{2}{c}{$\kappa=1$}& \\ \hline\hline
$C $    &     TDLR &      DVR   &   TDLR    & DVR     &\\ \hline
0       &    1.9995  & 2.0000   &  1.9996   & 2.0000  &\\
50      &    1.9998  & 2.0000   &  1.9998   & 2.0000  &\\
100     &    1.9936  & 2.0000   &  1.9979   & 2.0000  &\\
250     &    1.9977  & 2.0000   &  1.9980   & 2.0000  &\\
500     &    1.9978  & 2.0000   &  1.9978   & 2.0000  &\\
1000    &    1.9978  & 2.0000   &  1.9977   & 2.0000  &\\
\hline
\end{tabular}
\caption{Comparison of TDLR and DVR results for the
excitation frequencies $\omega_j$, for the mode
$q_{r}=1,q_{\theta}=0$  with $\kappa=0,1$.} 
\label{qr1qt0}
\end{table}

\begin{table}
\begin{tabular}{|c|c|c|}
\hline
 $C$      &   TDLR      &   DVR \\
\hline
    0     &     2.0000  & 2.0000\\
   50     &     1.5569  & 1.5569\\
  100     &     1.5003  & 1.5002\\
  250     &     1.4561  & 1.4561\\
  500     &     1.4380  & 1.4380\\
 1000     &     1.4277  & 1.4276\\
 $\infty$ &     1.4142  & 1.4142\\
\hline

\end{tabular}
\caption{Comparison of TDLR and DVR results for the quadrupole
excitation $q_{r}=0,q_{\theta}=\pm2$ mode with condensate $\kappa=0$.} 
\label{quad1}
\end{table}

\begin{table}
\begin{tabular}{|c|c|c|c|cc|}
\hline
 & \multicolumn{2}{c}{$q_{\theta}=-2$}&\multicolumn{2}{c}{$q_{\theta}=2$}& \\ \hline\hline
$ C$  &   TDLR & DVR &   TDLR & DVR &\\
\hline
    0&  0.0000 &    0.0000  &   2.0000 & 2.0000  &\\
   50&  0.8333 &    0.8331  &   1.8647 & 1.8649  &\\
  100&  1.0299 &    1.0300  &   1.7752 & 1.7748  &\\
  250&  1.1876 &    1.1871  &   1.6590 & 1.6594  &\\
  500&  1.2565 &    1.2560  &   1.5902 & 1.5906  &\\
 1000&  1.3027 &    1.3022  &   1.5388 & 1.5393  &\\
\hline
\end{tabular}
\caption{Comparison of TDLR and DVR results for quadrupole
excitation $q_{r}=0,q_{\theta}=\pm2$ for a condensate with
circulation $\kappa=1$.} 
\label{quadcomplz1m}
\end{table}

 The dilute system of $N$ bosonic atoms mass $m$
is trapped by external fields $V_{\rm ext}(\bm{r})$ and interacts
weakly through the  two-body $V(\bm{r},\bm{r}')$. If the
external field is  the static trapping potential, the Hamiltonian
for the system can be written
\begin{eqnarray}
\hat{H}' &=& \int d \bm{r}\ \hat{\Psi}^{\dag}( \bm{r}) H_0
\hat{\Psi}(\bf {r})  \nonumber \\ &+& {\textstyle{\frac{1}{2}}}
\int\!\! \int d \bm{r} d \bm{r}'
\hat{\Psi}^{\dag}(\bm{r})\hat{\Psi}^{\dag}(\bm {r}')
V(\bm{r},\bm{r}') \hat{\Psi}(\bm{r}')\hat{\Psi}(\bm{r}) \\
\nonumber
\end{eqnarray}
with  $H_0=-\frac{\hbar^{2}}{2m}\nabla^{2} + V_{\rm trap}(\bm
{r})-\mu$,  with chemical potential $\mu$. The interactions can be
represented perturbatively by the pseudopotential $V(\bm {r},\bm
{r}')=U_{0} \delta (\bm{r}
 - \bm{r}')$ where the interaction is proportional to the $s$-wave scattering
length $a$, $U_{0}= 4\pi\hbar^{2}a/m$. Bogoliubov's approximation,
valid at low temperatures, treats the condensate semiclassically
while quantizing the excitations so that:
\begin{equation}
\hat{\Psi}(\bm {r}) = \sqrt{N}\ \phi(\bm {r}) + \hat{\psi}(\bm {r})
\end{equation}
The condensate wavefunction satisfies the Gross-Pitaevskii
equation:
\begin{equation}
H_{0}(\bm{r})\phi(\bm{r}) +NU_{0}|{\phi}(\bm{r})|^{2} \phi(\bm{r})=0
\end{equation}
Then expanding $\hat{\psi}$ in  modes
\begin{equation}
\hat{\psi}({\bm r}) =
 \sum_{j} {\left[u_{j}(\bm {r})\hat{\alpha}_{j} +
 v_{j}^{\ast}(\bm {r})\hat{\alpha}^{\dag}_{j} \right]}
\end{equation}
The eigenvalue problem reduces to the Bogoliubov-deGennes
equations in the form:
\begin{subequations}
\begin{eqnarray}
{\cal{L}}u_{j}(\bm{r}) + NU_0 \phi^2 v_j(\bm{r}) &=&
\hbar \omega_{j}u_{j}(\bm {r})\label{bdg1}
\\
{\cal{L}}v_{j}(\bm {r}) + NU_0 \phi^{\ast}{}^{2}u_{j}(\bm {r}) &=&
-\hbar \omega_{j}v_{j}(\bm {r})\label{bdg2}
\end{eqnarray}
\end{subequations}

where ${\cal{L}} = H_{0}(\bm {r}) +2NU_{0}|{\phi}(\bm {r})|^{2}$.
Time-reversal symmetry is reflected in the fact that every set of
solutions $\{E_j,u_{j},v_{j}\}$ has a corresponding set
$\{-E_j,u^{\ast}_{j},v^{\ast}_{j}\}$. For $E_{j}> 0$ the functions
$u_j,v_j$ are orthogonal with normalization
\begin{equation}
\int d \bm {r} \left( |u_j(\bm {r})|^{2} - |v_j(\bm {r})
|^{2}\right) = 1
\end{equation}

\begin{figure}
\centering
\includegraphics[width=6cm]{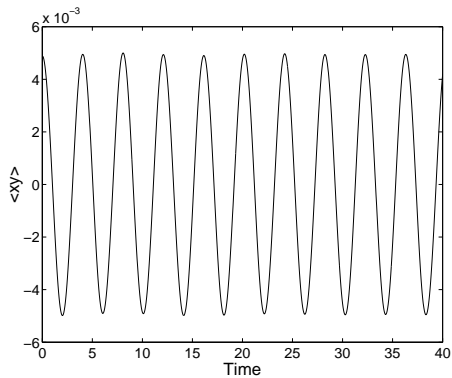}
\caption{ Quadrupole moment evolution $f_{xy}(t)$ for low-lying
excitations of a condensate with $\kappa=0$.}
 \label{fig:xylz0}
\end{figure}

\begin{figure}
\centering
\includegraphics[width=6cm]{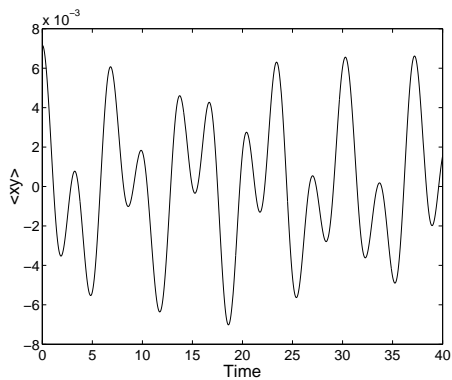}
\caption{ Quadrupole moment evolution $f_{xy}(t)$ for low-lying
excitations of a condensate with $\kappa=1$. The figure displays
interference between the modes $q_{\theta}=\pm 2$}
\label{fig:xylz1}
\end{figure}

\begin{figure}
\centering
\includegraphics[width=6cm]{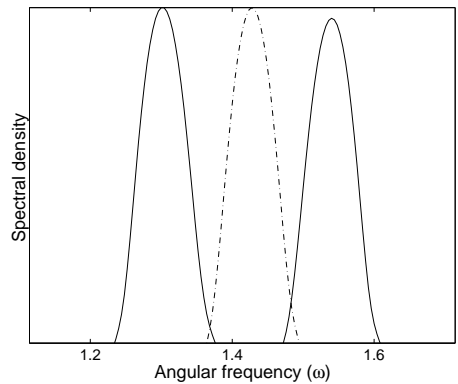}
\caption{Power spectral density of quadrupole moments $\langle
xy\rangle$  versus angular frequency. The vertical
scale is logarithmic and ranges from 0.1 to 1.0.
Results for $\kappa=0$, dashed
line, $q_{\theta}=\pm 2 $ and $\kappa=1$, solid lines,
$q_{\theta}=\pm 2$ for interaction strength  $C=1000$.}
\label{fig:pow_C1000}
\end{figure}

In our model the trap is represented by an asymmetric harmonic
well:
$$
V_{\rm trap} (\bm {r})= {\textstyle{1 \over 2}}m \omega_0^2(e_x^2
x^2+e_y^2 y^2 + e_z^2 z^2 )
$$
where $e_{x,y,z}$ are  restoring force strengths, and $\omega_0$
the natural angular frequency. Numerical calculations are carried
out in scaled dimensionless units. Length, time and energy are given in
units : $\left(\hbar/2m\omega_0 \right)^\frac{1}{2}$,
$\omega_0^{-1}$ and $\hbar \omega_0$, respectively.

When particle interactions are weak, corresponding to the ideal
gas limit, the energies for an
asymmetric trap are given by the oscillator formula
$$
E_n = \left( n_x  +{\textstyle{1 \over 2}} \right) e_x+
\left( n_y  +{\textstyle{1 \over 2}} \right) e_y+
\left( n_z +{\textstyle{1 \over 2}}\right) e_z
$$
with $n_x,n_y,n_z=0,1,2,\dots$. For a 2D symmetric oscillator
($e_z=0,e_x=e_y=1$) the cylindrical quantum numbers ($q_r,q_{\theta}$)
can be used:
$$
E_n = n_x+n_y+1 = 2q_r+\vert q_{\theta} \vert+1
$$
with $q_r=0,1,2,\dots$ and $q_{\theta}=0,\pm 1,\pm 2,\dots $ so that
the excitation frequencies above the ground state ($q_r=0,q_{\theta}=\kappa$)
are:
$\omega_j = 2q_r+|q_{\theta}+\kappa| -\kappa$.

\subsection{Excitations of vortex states}

 A precise test is provided by low-dimensional
models  for which highly accurate results can be obtained. It also
allows the study of excitations in the presence of vortices. For a
single vortex line with circulation $\kappa$ along the axis of a
cylindrically symmetric trap $e_x=e_y=1,e_z=0$:
\begin{equation}
\phi(r,\theta) = \tilde{\phi}(r)e^{i\kappa\theta}
\end{equation}

\begin{eqnarray}
 -\left( \frac{\partial^{2}}{\partial r^{2}} +  \frac{1}{r}
\frac{\partial}{\partial r} - \frac{\kappa^{2}}{r^{2}}
\right)\tilde{\phi}(r)
&+& {\textstyle{\frac{1}{4}}} r^{2}\tilde{\phi}(r) \label{gpe}   \\
&+& C |\tilde{\phi}|^{2}\tilde{\phi}(r)
  =  \mu \tilde{\phi}(r) \nonumber
\end{eqnarray}
where $C = 8\pi a N_a$ with $N_{a}$ the
linear density of atoms in the $\hat{z}$ direction.
The excitations can be written:
\begin{eqnarray}
u_{j}(r,\theta) = \tilde{u}_{q_{r}}(r) e^{i(q_{\theta} +
\kappa)\theta} \\  \nonumber
 v_{j}(r,\theta) =
\tilde{v}_{q_{r}}(r) e^{i(q_{\theta} - \kappa)\theta}
\end{eqnarray}
where $j=\{q_{r},q_{\theta}\}$.
The eigensystem equations (\ref{bdg1},\ref{bdg2}) become
\begin{eqnarray}
\left[- \frac{\partial^{2}}{\partial r^{2}} -\frac{1}{r}
\frac{\partial}{\partial r} + \right.
& \left. \!\!\!\! {\displaystyle \frac{(q_{\theta}+\kappa)^{2}}{r^{2}}}
+ {\textstyle {\frac{1}{4}}} r^2 - \mu  + 2 C
|\tilde{\phi}|^{2}\right]
\tilde{u}_{q_{r}}(r) \nonumber \\
&+ C \tilde{\phi}^{2}\tilde{v}_{q_{r}}(r) =
\omega_{q_{r}}\tilde{u}_{q_{r}}(r)
\label{bdgone}
\end{eqnarray}
\begin{eqnarray}
\left[- \frac{\partial^{2}}{\partial r^{2}} -\frac{1}{r}
\frac{\partial}{\partial r} + \right.
&\left. \!\!\!\! {\displaystyle \frac{(q_{\theta}-\kappa)^{2}}{r^{2}}}
+ {\textstyle {\frac{1}{4}}} r^2 - \mu  + 2 C
|\tilde{\phi}|^{2}\right]
\tilde{v}_{q_{r}}(r) \nonumber \\
&+ C \tilde{\phi}^{2}\tilde{u}_{q_{r}}(r) =
-\omega_{q_{r}}\tilde{v}_{q_{r}}(r)
\label{bdgtwo}
\end{eqnarray}

\subsection{Discete variable representation}

Equations (\ref{gpe}),(\ref{bdgone}) and (\ref{bdgtwo}) are solved
using the discrete variable representation (DVR) \cite{bay86}. The
condensate and excitation functions are represented by a Lagrange
mesh $\tilde{\phi},~\tilde{u}_{q_{r}},~\tilde{v}_{q_{r}}$.
\begin{equation}
\tilde{\phi}(\bm{r}) = \sum_{i=1}^{N} \tilde{\phi} (r_{i})
f_{i}(r)
\end{equation}
\begin{equation}
\tilde{u}_{q_{r}}(\bm{r}) = \sum_{i=1}^{N} \tilde{u}_{q_{r}} (r_{i})
f_{i}(r)
\end{equation}
\begin{equation}
\tilde{v}_{q_{r}}(\bm{r}) = \sum_{i=1}^{N} \tilde{v}_{q_{r}} (r_{i})
f_{i}(r)
\end{equation}
where $f_{i}(r)$ are Lagrange interpolating functions constructed
from a set of orthonormal functions
\begin{equation}
\chi_{k}(r) = h_{k}^{-\frac{1}{2}}w(r)^{\frac{1}{2}} p_{k}(r)
\end{equation}
such that
\begin{equation}
 f_{i}(r) = \sum_{k=0}^{N-1} \chi_{k}^{\ast} (r_{i})  \chi_{k}(r)
\end{equation}
We employ generalized Laguerre polynomials \cite{abr65} $p_k(r)=
L_{k}^{\alpha}(r) =  ((\alpha+k)!/ k!\alpha !) {}_1F_1(-k;\alpha+1
;r)$ associated with the weight function $w(r)=r^{\alpha}e^{-r}$
and normalization factor $h_{k} = (k+\alpha)!/k!$. The
representation of the centrifugal energy terms of equations
(\ref{bdgone}),(\ref{bdgtwo}) can be marginally improved by using
different values of $\alpha$ \cite{bay86}. Different grids for
each angular momentum state  requires the use of interpolation to
connect different states for $\kappa \neq 0$. However using a
common grid  is more practical. The DVR results presented
correspond to $k=N=50$ and the common value $\alpha=1$ so that:
\begin{equation}
f_{i}(r) = \frac{1}{\chi_{N}^{'}(r_{i})}
\frac{\chi_{N}(r)}{r-r_{i}}
\end{equation}
where the mesh points $r_{i}$ are the $N$ zeros of
$L_N^{\alpha}(r)$. Applied to the coupled equations
(\ref{gpe}),(\ref{bdgone}),(\ref{bdgtwo}) this method reduces the
problem to that of solving a system of nonlinear equations for the
condensate and a linear eigenvalue problem for the excitations.
The set of ($N+1$) nonlinear equations is efficiently solved using
Newton's method where $\mu$ is an unknown and normalization of
$\tilde{\phi}$ is imposed. As the functions
$\tilde{\phi},\tilde{u},\tilde{v}$ are to be represented on the
same grid points, $r_i$, an adequate
coverage of points is required. A scaling factor $h$
\cite{bay86} is used to contract the natural mesh so that it
extends to  $\sim 1.5r_{\rm TF}$ where $r_{\rm TF}
\approx 2(C/2\pi)^{1/4}$ is the Thomas-Fermi radius. Increasing
the number of mesh points beyond $N=50$  does not  improve  the
accuracy beyond the figures quoted. While the matrices are dense
 they give results far more accurate than  finite difference methods using
similar sized matrices. We used a standard library routine for
solution of the eigenvalue problem  \cite{nag00}.

\section{Time-dependent linear response equations }\label{sec:tdlr}

The mean-field approximation to the field operator is
$\psi(\bm{r},t) \approx  \langle \Psi(\bm{r},t) \rangle \approx
\langle \Psi^{\dag}(\bm{r},t) \rangle $. The angular brackets
denote averaging with respect to highly-occupied  ($N \gg 1$)
condensate number states. The evolution equation for the mean
field is the time-dependent Gross-Pitaevskii equation
\cite{pin60}:
\begin{equation}
i\hbar \frac{\partial}{\partial t} \psi(\mbox{\boldmath $r$},t)
 =  \left[-\frac{\hbar^2}{2m}\nabla^2 +
V_{\rm ext}(\mbox{\boldmath $r$},t) + N U_0
\vert\psi(\mbox{\boldmath $r$},t)\vert^2\right]
\psi(\mbox{\boldmath $r$},t) \label{eq:nlse1}
\end{equation}
The  external potential includes the trapping potential and a
small time-dependent perturbation: $ V_{\rm ext}(\bm{r},t)=V_{\rm
trap}(\bm{r})+ V_{\rm pert}(\bm{r},t) $. The system responds to
this perturbation  by populating  modes selected by the symmetry
of the perturbation:
\begin{eqnarray}
\psi(\bm{r},t)  & \approx &   e^{-i\mu t} \phi ({\bm r})  \\
 &  + & \sum_{j} \left[ a_j(t) u_j(\bm{r}) e^{-i\omega_j
t-i\mu t}+b_j (t) v_j (\bm{r})e^{i\omega t-i\mu t}\right] \nonumber
\end{eqnarray}
The oscillations  of the system provides data for the
frequencies: $\omega_j$. This replaces the nonlinear
coupled equation eigenvalue problem by an initial value problem
followed by spectral analysis. Alternatively, we can introduce a
perturbed initial state $\Psi({\bm r},0)$ seeded with the spatial
form to overlap with the symmetry of the desired excited state.
This is equivalent to a sudden perturbation of the system.  
To study low-energy excitations the ground state, $
\phi ({\bm r})$, is found by propagation of equation (\ref{eq:nlse1}) 
in imaginary time \cite{win00} using  a trial wavefunction.  Implementing  the
split-operator FFT method on a parallel computer using between
 1 and 16 processors. The method  is extremely well-suited for
nonlinear evolution equations \cite{tah84}. In imaginary time 
the highly-excited
modes are exponentially suppressed, and while it is not necessary to
eliminate these modes  before introducing  
the perturbation, a pure ground state is more
efficient for  spectral analysis.

The dipole modes were excited by a sudden but small displacement 
of the trap center. The subsequent 
center-of-mass oscillations are known 
exactly ($\omega=e_x,e_y,e_z$) and
should provide precise benchmarks for the numerical method 
employed. 
Breathing (monopole) modes were excited
selectively, and in line with experimental methods \cite{jin96}, by
adjusting the force constants to compress or expand the condensate
$e_{x,y} \rightarrow 1 \pm \varepsilon$. The low-lying  quadrupole
modes are accessible by seeding the initial state with a term 
of the appropriate symmetry. For example, adjusting the true ground state
to
$\Psi({\bm r},0)=\phi({\bm r})(1+\varepsilon xy)$, and following
the evolution of  $\Psi ({\bm r},t)$  under the 
Gross-Pitaevskii equation, the beats of the $xy$ quadrupole spectrum
 are evident. While single modes may be selected by adiabatic perturbations
 of selected frequencies, sudden and strong perturbations can be used 
 to gain data on higher excited modes.
 A significant
advantage of the method is that a variety of perturbations of
{\sl different} symmetries can be used simultaneously to obtain the
excitations by a {\sl single} evolution of the nonlinear equations. 

The wave packet method combined with spectral method was pioneered
by Feit and Fleck \cite{fei82} and has been used widely 
\cite{lig85,erm96,kos88}. The response of the system is measured
by the moment:
\begin{equation}
f_A(t)= \int d\bm{r}  \Psi^*(\bm{r},t) A(\bm{r})  \Psi(\bm{r},t)
\end{equation}
The excitation frequencies emerge from the Fourier transform of
$f_{A}(t)$. The finite sampling time means that the power spectral
density signal can be improved by windowing. The windowing
function we use is a $\beta$-valued Kaiser function chosen to
minimize the background. The local maximum of the power spectrum
is then used to identify the frequencies. The density profiles of
the quasiparticles can be extracted by further spectral 
analysis. Once the appropriate $\omega_j$ is known the 
Fourier transform:
$g^{\pm}_j(\bm{r})=T^{-1}\int_0^T e^{ i(\mu \pm \omega_j) t} \Psi(\bm{r},t) \ dt\ $
isolates the corresponding amplitude.
For both the imaginary time and real time propagation we used the
same split-operator scheme. In the calculations we used a grid of
64 points in each dimension.
To create the ground state we used a variable time step along with
extrapolation to increase the speed of calculation and control the
accuracy of the ground state. To obtain an accurate spectral
resolution a total propagation time  of  $\sim$30 times the trap
period is necessary. In addition, the signal was processed with a
windowing mask. Alternatively if the perturbation is small and
introduced slowly, on the timescale of several trap periods, only
a very small number of modes will be excited. The correlation
$f_{A}(t)$ can be fitted to a Fourier expansion in just a few
frequencies. This has been done for all the frequencies presented
in this paper and gives equivalent accuracy while requiring  only
3 trap periods.

\subsection{Classical limit}
When interparticle forces dominate the quantum pressure ($C
\rightarrow \infty$) and the hydrodynamic limit is reached. The
equilibrium condensate density, $\rho\equiv \vert \psi \vert^2$,
for $\kappa=0$ corresponds to the  Thomas-Fermi distribution:
\begin{equation}
\rho_0 (  \bm{r} )= \left( \mu - V(\bm{r}) \right)/C
\end{equation}
Acoustic modes of excitation \cite{pit80}
$\rho(\bm{r},t)=\rho_0(\bm{r}) + \rho' (\bm{r})\  e^{-i\omega t}$
are determined by the equation \cite{str96}
\begin{equation}\label{acoustic}
\omega^2 \rho' = -2C  \nabla \cdot \left( \rho_0 \nabla  \rho'
\right)
\end{equation}
For the 2D cylindrical trap  ($e_x=e_y=1, e_z=0$) this gives
($C \rightarrow \infty $):
\begin{equation}
\omega_c^2= 2 q_r (q_r+1) +q_{\theta} (2 q_r+1).
\label{hydro}
\end{equation}
where the radial and angular quantum numbers ($q_r,q_{\theta}$)
are as defined above.

For a 3D asymmetric trap ($e_x \neq e_y \neq e_z $) the lowest
quadrupole modes of scissors motion \cite{str96,gue99,mar00} that
is of the form $\rho' \propto xy,yz,xz$ are given by the
solutions:
 $\omega^2_{xy}=e_x^2+e_y^2,\ \omega^2_{yz}=e_y^2+e_z^2,
 \  {\rm and} \ \omega^2_{zx}=e_z^2+e_x^2 $.

\begin{table}
\begin{tabular}{|c|c|c|c|}
\hline
$C$       &  $\omega_{xy}$  & $\omega_{xz}$ & $\omega_{yz}$ \\
\hline
    0     & 3.0005&    4.0003&    5.0002\\
   10     & 2.8059&    3.8041&    4.7938\\
   50     & 2.5439&    3.5536&    4.4612\\
  100     & 2.4432&    3.4228&    4.2521\\
  250     & 2.3521&    3.3134&    4.0254\\
  500     & 2.3101&    3.2609&    3.8951\\
 1000     & 2.2821&    3.2257&    3.7993\\
$ \infty$ & 2.2361&    3.1623&    3.6056\\
 \hline
\end{tabular}
\caption{Lowest-energy quadrupole modes for an asymmetric trap
with $e_x=1,e_y=2$ and $e_z=3$ as a function of interaction
strength $C$. Results for $C =\infty$ correspond to the
hydrodynamic limit (\ref{hydro})} 
\label{asymm}
\end{table}

\section{Results}
\label{sec:res}
 In this article we compare the time-dependent (TDLR) and time-independent
 methods (DVR) for excitation in 2D traps. For the DVR calculations
 we used $N=30$ and $N=50$ radial mesh points with a scaling factor $h=0.08$.
 The results presented, using $N=50$,
  are converged to at least 6
 decimal places. For reference, the values of the chemical
 potential are presented in Table \ref{chempot}. The TDLR calculations
 were performed on grids of $64\times 64 \times 64$ points.
 Convergence and accuracy can be calibrated by comparison with 
 exact results. For example, the 
  2D dipole ($q_r=0,q_{\theta}=1$) mode is a feature of the trap and 
  independent of the particle interactions ($C$) and
circulation $\kappa$, so that $\omega_j=1$. The DVR results for
 this mode are accurate
to at least six decimal places, while the TDLR results are within
0.3\% of the exact values over the range $0 \leq C \leq 1000$.
 Table
\ref{qr1qt0} shows the results
for the lowest breathing mode ($q_r=1,q_{\theta}=0$) for $\kappa=0$
and $\kappa=1$. Again, in 2D this mode is not sensitive to particle
interactions. TDLR results are stable and consistent for this mode
as $C$ increases, slightly underestimating the frequency in 
the fourth significant figure. The loss of precision in the TDLR results 
for high $C$ is of the same order as the 
errors in the chemical potential results. 

The quadrupole modes are more sensitive to changes in  $C$ and 
also to the condensate circulation. In figure \ref{fig:xylz0} the 
correlation $f_{xy}(t)$ is shown for $C=1000$. The smooth 
sinusoidal curve translates to the Fourier
transform spectral density shown as the dashed line in figure 
\ref{fig:pow_C1000}. The spectral density shows a noise-free 
profile with a well-defined peak at the value
$\omega= 1.4277$. As $C\rightarrow +\infty$, $\omega \rightarrow \sqrt{2}$.
In Table \ref{quad1} results are presented for the mode:
$\kappa=0,q_r=0,q_{\theta}=\pm2$ . The agreement between
the time-dependent and time independent results is extremely good 
over the entire range of $C$. Given that the dipole mode showed the
TDLR accuracy is within four figures, the accord to five figures
 for $C=250$ and $C=500$ is fortuitous. 

When condensate circulation is present the quadrupole circulation can 
flow with or against the background and the $q_{\theta}=\pm 2$ modes
split. This is evident in the beats arising in the $f_{xy}(t)$ correlation
figure \ref{fig:xylz1} and the 
corresponding spectrum (figure \ref{fig:pow_C1000}). The agreement with DVR 
results, indicated in Table \ref{quadcomplz1m} is excellent. 
Since  the TDLR method is based upon a cartesian grid, it is readily
used for arbitrary geometries of the trap. For example, the quadrupole
modes $xy,yz,zx$ of an asymmetric 3D ellipsoidal trap can found in 
the same way. In this case we used a mixed perturbation of all three
symmetries and then frequency analyzed the correlations of each moment. 
The results are presented in Table \ref{asymm}. The data tends to the
correct high $C$ limit uniformly and accurately.

\section{Conclusion}\label{sec:conc}
We have compared  methods for calculation of the excitation
frequencies of a condensate at zero temperature. The method
based on linear response theory is extremely efficient
and successful at
producing accurate results and can be tailored to the
experimental methods used to create the excitations
of interest. The method is straightforward to implement,
yields reliable results, and can be computed cheaply and
efficiently. The versatility of the method
extends to complex trap geometries and condensate topologies. Thus
it is suitable for the study of excitations and sound propagation
in traps where the condensate contains soliton and vortex structures 
\cite{abo01}.

We are very grateful for the support of Bergen Computational
Physics Laboratory in the framework of the European Community -
Access to Research Infrastructure action of the Improving Human
Potential Programme.


\end{document}